# MoSe$_2$ and WSe$_2$ shell morphology control via temperature optimization during two-step growth of ZnSe-based core-shell nanowires

*Luize Dipane[1], Liora Kotlara[1], Viktors Vibornijs[1], Katrina Laganovska[1], Aleksejs Zolotarjovs[1], Eriks Dipans[1], Jevgenijs Gabrusenoks[1], Boris Polyakov[1], Edgars Butanovs[1]*

[1]Institute of Solid State Physics, University of Latvia, Kengaraga street 8, Riga, Latvia, LV-1063

**Abstract**

Achieving uniform and controlled transition metal dichalcogenide (TMD) shell growth on nanowires (NWs) remains a key challenge, limiting the development of high-quality core-shell heterostructures for optoelectronic and photocatalytic applications. In this work, the fabrication of ZnSe-MoSe$_2$ and ZnSe-WSe$_2$ core-shell NWs was successfully demonstrated. ZnSe NWs were grown via the vapor-liquid-solid growth mechanism, while TMD (MoSe$_2$ or WSe$_2$) shells were formed through a two-step process of sacrificial oxide layer deposition via magnetron sputtering followed by selenization process in a chemical vapor transport reactor. As-grown nanostructures were characterized using X-ray diffraction, transmission electron microscopy, X-ray photoelectron spectroscopy, Raman spectroscopy and photoluminescence spectroscopy. It was observed that the TMD shell morphology can be controlled through the selenization process temperature optimization, which arises due to different growth mechanisms discussed here. The studied trends could be further extended to other semiconductor NW and TMD core-shell heterostructure growth, offering promising avenues for advanced nanoscale applications.

# 1. Introduction

Over the past two decades, nanostructured materials, especially semiconductor heterostructures, have been major contributions to advance nanotechnology [1,2]. There has been a particularly growing interest in nanowires (NWs) research [3,4], as semiconductor NWs have emerged as versatile building blocks for nanoscale devices due to their high aspect ratios and tunable electronic properties [5]. Extensive studies on materials such as Si, GaAs, and ZnO have demonstrated their use in applications ranging from sensors to photonic devices [6].

Zinc selenide (ZnSe) is II-VI semiconductor with a wide bandgap of approximately 2.7 eV, which makes it suitable for applications in blue light-emitting devices and infrared optics [7]. ZnSe NWs exhibit unique optical and electronic properties influenced by quantum confinement effects, which have been widely investigated [8,9], positioning them as promising candidates for optoelectronic applications, including photodetectors [10] and light-emitting diodes [11].

Transition metal dichalcogenides (TMDs) exhibit a distinctive electronic structure [12] and a tunable optoelectronic response [13], which make them suitable for applications in transistors [14], photodetectors [15], and other optoelectronic components [16]. Tungsten diselenide ($WSe_2$) and molybdenum diselenide ($MoSe_2$) are TMD materials with thickness-dependent bandgaps and high stability, making them promising for applications in flexible electronics, tunnelling transistors, optoelectronics, and energy storage devices [16–18]. Due to the favourable band alignment between ZnSe and $WSe_2$ [19], as well as ZnSe and $MoSe_2$ [20–22], heterostructures composed of these materials can facilitate efficient charge separation and directional carrier transport, which is beneficial for applications such as photocatalysis and photoelectrochemical energy conversion.

Core-shell heterostructures are of particular interest due to their ability to combine the individual properties of both core and shell materials, enabling tailored optoelectronic characteristics and enhanced stability [23,24]. Several strategies have been developed to integrate TMDs with NWs, including direct chemical vapor deposition growth [25], atomic layer deposition [26], and post-growth conversion of oxide layers [27]. Sacrificial oxide film selenization and sulfurization, in particular, have proven effective for synthesizing high-quality TMD coatings by converting pre-deposited metal or metal oxide layer into corresponding chalcogenides under controlled atmospheres [28]. This approach allows for precise control over thickness, crystallinity, and phase composition, which are critical parameters for ensuring reliable device performance. Despite the clear advantages, achieving controlled and uniform TMD shell growth on semiconductor NWs requires precise synthesis techniques and careful optimization of processing parameters [29].

In this work, ZnSe-WSe$_2$ and ZnSe-MoSe$_2$ core-shell NWs were fabricated by depositing a corresponding precursor oxide layer onto ZnSe NWs via DC magnetron sputtering, followed by selenization in a three-zone horizontal furnace to form the TMD shell. A variety of characterization techniques were employed to investigate the influence of selenization temperature on the morphology and crystalline quality of the TMD shells. Control over these parameters could enable full optimization of the core-shell heterostructure properties required for applications in photocatalysis and optoelectronics.

## 2. Materials and Methods

The ZnSe NWs were synthesized using the chemical vapor deposition (CVD) method under atmospheric pressure. The process was performed in a horizontal three-zone quartz tube reactor (Carbolite Gradient Tube Furnace TG3-12-60-600). A ceramic boat containing 0.06 g of ZnSe powder (99.99%, Aldrich) was placed in the first zone at 1000 °C, while SiO$_2$/Si(100) wafers (SemiconductorWafer, Inc.) coated with Au nanoparticles (NPs, ThermoScientific, 40 nm diameter) were positioned in the second zone at 750 °C. The Au nanoparticles acted as catalysts for the vapor-liquid-solid (VLS) growth of NWs. The system was maintained under the mixture of Ar/H$_2$ (5%) gas flow at 200 standard cubic centimetres per minute (sccm) and N$_2$ at 50 sccm, then heated until the target temperatures (1000°C/750°C/650°C) were reached in all zones. The temperature was then held for 20 minutes to facilitate NW growth before allowing the system to cool naturally to room temperature. See Fig.S1 for the characterization results of the as-grown ZnSe NWs.

Following the synthesis of ZnSe NWs, a subsequent deposition of molybdenum oxide (MoO$_3$) and tungsten oxide (WO$_3$) was carried out on the ZnSe NWs. This deposition process was performed using a reactive direct-current (DC) magnetron sputtering technique, ensuring precise and controlled film growth. In this method, metallic tungsten (W, 99,95%) and molybdenum (Mo, 99,95%) targets, respectively, were utilized as source materials in a precisely controlled mixed argon (Ar, 99.999999%) and oxygen (O$_2$ 99.99999%) gases, with an Ar:O$_2$ ratio of 3:2. The sputtering power was maintained at 300 W to achieve optimal deposition conditions and the pressure during the sputtering process was maintained at 5 mTorr. To achieve the desired final material properties, the thickness of the deposited thin films was meticulously optimized for each oxide, the deposited layers are amorphous MoO$_3$ and WO$_3$, respectively (see Fig.S2 and S3). Specifically, a thin film of 15 nm thickness was deposited for MoO$_3$, while WO$_3$ was deposited to a thickness of 20 nm. The ability to finely tune the thickness and uniformity of these oxide films is important, as it directly influences the structural properties of the resulting ZnSe-based heterostructures structures.

For the synthesis of ZnSe-MoSe$_2$ and ZnSe-WSe$_2$ core-shell NWs, a three-zone horizontal quartz tube reactor was used. The precursor samples, ZnSe-MoO$_3$ or ZnSe-WO$_3$, as well as thin film samples for reference, were positioned in the central zone of the reactor, while selenium powder (>99%, Alfa Aesar) was placed in the first zone and heated to 400°C. The reaction was conducted under a controlled Ar/H$_2$ (5%) atmosphere at a constant flow rate of 60 sccm. To achieve optimal material characteristics, the temperature of the central zone and the reaction duration were systematically optimized for each composition. The synthesis was performed at varying temperatures to investigate their impact on shell growth, surface morphology, and phase composition - ZnSe-MoSe$_2$ NWs at 600 and 650 °C, and ZnSe-WSe$_2$ NWs at 700 and 750 °C. Above these maximum temperatures typically no coating was remaining, while below these minimum temperatures conversion of the oxide precursor film was not fully complete. For both compounds, an optimum reaction time of 20 minutes was determined to ensure uniform shell formation and high-quality phase composition. The integrity of ZnSe core NWs is typically maintained to at least 800 °C (see Fig.S4)

The morphology of the as-synthesized NWs was characterized by transmission electron microscopy (TEM, Tecnai GF20, FEI) operated at a 200 kV accelerating voltage for examining their inner crystalline structure. X-ray diffraction (XRD) analysis using Rigaku MiniFlex 600 powder diffractometer with Bragg-Brentano θ-2θ geometry and a 600 W Cu anode X-ray tube (Cu Kα line, λ = 1.5406 Å) was performed to analyze the phase composition of the NWs. The chemical composition of the NWs and thin films was studied with X-ray photoelectron spectroscopy (XPS) measurements performed using ESCALAB Xi spectrometer (ThermoFisher). Al Kα X-ray tube with the energy of 1486 eV was used as an excitation source, the size of the analysed sample area was 650 μm x 100 μm and the angle between the analyser and the sample surface was 90°. An electron gun was used to perform charge compensation; however, no sputter-cleaning was performed prior the measurements. The base pressure during the spectra acquisition was better than $10^{-5}$ Pa.

Micro-Raman spectroscopy was conducted using a TriVista 777 confocal Raman system (Princeton Instruments) with a 532 nm continuous-wave single-frequency laser source. An upright Olympus microscope equipped with an Olympus UISe MPlanN 100x/0.90 objective (light spot size around 0.5-1 μm) was utilized for measurements on individual free-standing NW. Photoluminescence (PL) measurements were carried out at room temperature and lower temperatures using the 4th harmonic (266 nm, 4.66 eV) of a YAG laser FQSS266 (CryLas GmbH). PL spectra were acquired with a Horiba iHR320 spectrograph coupled to a CCD camera (Andor DV420A-BU2).

## 3. Results

TEM analysis confirmed the formation of core-shell structure (see Fig. 1 and Fig.S5). The $WSe_2$ shell formed on ZnSe core-shell NWs selenized at 700 °C, Fig. 1 (a,b), was approximately 2 nm thick, consisting of continuous three layers. The interlayer distance was measured to be approximately 6.8 Å for the (002) orientation, consistent with values mentioned in literature [30]. However, for ZnSe-$WSe_2$ NWs selenized at 750 °C, the shell layer was discontinuous, as shown in Fig. 1 (c,d), no continuous layer was observed on the NW surface. Instead, large protruding $WSe_2$ crystals, 20-50 nm in size, were formed, as confirmed by an interlayer distance of 6.8 Å corresponding to the (002) plane. For ZnSe-$MoSe_2$ NWs selenized at 600 °C Fig. 1 (e,f), the $MoSe_2$ shell was less than 2 nm thick and composed of two layers. The interlayer distance was measured to be 6.7 Å, corresponding to the (002) orientation [31]. For samples selenized at 650 °C Fig. 1 (g,h), similar to the behaviour observed in ZnSe-$WSe_2$ NWs at higher temperature, the NW surface was mostly covered with large $MoSe_2$ crystals, exhibiting the same interlayer distance of 6.7 Å (002). The interlayer distances within the ZnSe NWs, as shown in Fig. 1 (f), were measured to be approximately 6.5 Å and 3.5 Å, corresponding to the (002) and (100) planes, respectively, consistent with the wurtzite structure [32].

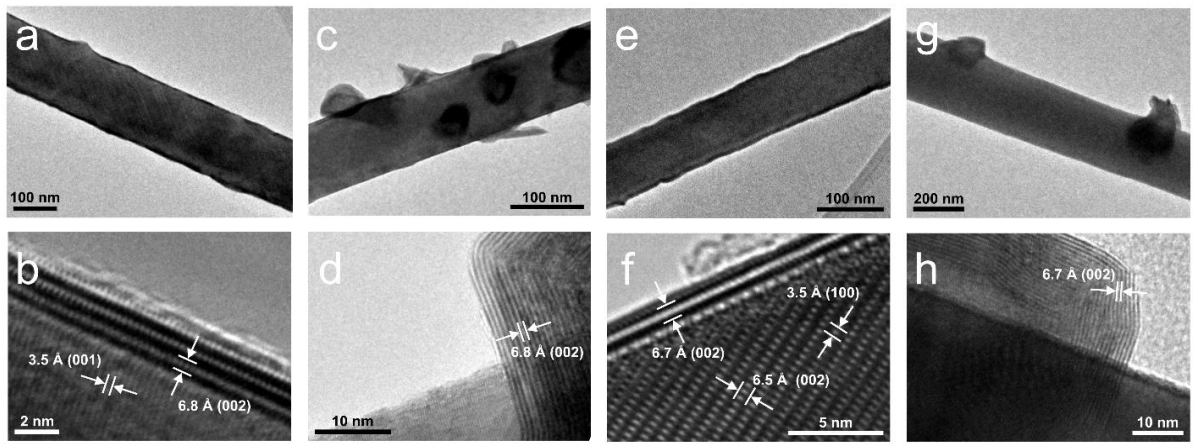

**Figure 1.** Transmission electron microscope images at different magnifications of (a,b) ZnSe-WSe$_2$ core-shell NWs selenized at 700°C and (c,d) 750°C. (e,f) ZnSe-MoSe$_2$ core-shell NWs selenized at 600°C and (g,h) 650°C. The insets display the measured atomic interlayer distance in the resulting phases.

XRD analysis Fig. 2 confirmed the formation of WSe$_2$ and MoSe$_2$ coating from the oxide precursor in the thin films and the core-shell NWs. The diffraction patterns showed distinct peaks corresponding to wurtzite ZnSe in the NW core (ICDD-PDF No: 00-015-0105), along with additional peaks attributed to MoSe$_2$ (ICDD-PDF No: 00-029-0914) and WSe$_2$ (ICDD-PDF No: 00-038–1388), indicating successful formation of the respective chalcogenide shells. No peaks corresponding to the precursor oxides were detected, and no quantitative differences were observed in the diffraction patterns across different selenization temperatures. The Bragg peak at 2θ ≈ 33° is attributed to the Si(100) substrate, corresponding to the forbidden Si(200) reflection.

XPS analysis was performed to verify the chemical states of the constituent elements in the NW and films (Fig.3 and Fig.S6). Survey spectra indicated the presence of the respective elements in the ZnSe, WSe$_2$ and MoSe$_2$ compounds, as well as carbon and oxygen from the organic surface contaminants. High-resolution spectra of Zn, W, Mo, and Se elements were acquired and calibrated relative to the adventitious C 1s peak at 284.8 eV. Zn 2p$_{3/2}$ peaks at 1021.9 eV (spin-orbit splitting $\Delta_{3/2-1/2}$ = 23 eV) were present in all pure and core-shell NW samples, while Se 3d$_{5/2}$ peak in the doublet for ZnSe compound was located at around 53.9 eV (spin-orbit splitting $\Delta_{5/2-3/2}$ = 0.86 eV) [33]. For WSe$_2$ thin film and ZnSe-WSe$_2$ core-shell NWs, W 4f$_{7/2}$ peak (spin-orbit splitting $\Delta_{7/2-5/2}$ = 2.17 eV) was found at around 32.4 eV with a W 5p$_{3/2}$ feature at around 38 eV [34], matching the WSe$_2$ chemical state; while for MoSe$_2$ thin film and ZnSe-MoSe$_2$ core-shell NWs, Mo 3d5/2 peak (spin-orbit splitting $\Delta_{5/2-3/2}$ = 3.15 eV) was present at 229.1 eV and attributed to the MoSe$_2$ compound [35]. Se 3d$_{5/2}$ peaks in both WSe$_2$ and MoSe$_2$ thin films were located at 54.7 eV [34,35], at a higher energy than in ZnSe spectrum. Consequently, two pairs of Se 3d doublets, one for ZnSe and one for diselenide compounds, with 0.4 – 0.5 eV chemical shift was distinguished in the ZnSe-WSe$_2$ and ZnSe-MoSe$_2$ selenium high-resolution scans. A feature at around 59 eV could possibly be attributed to a native surface oxide on the selenides [36].

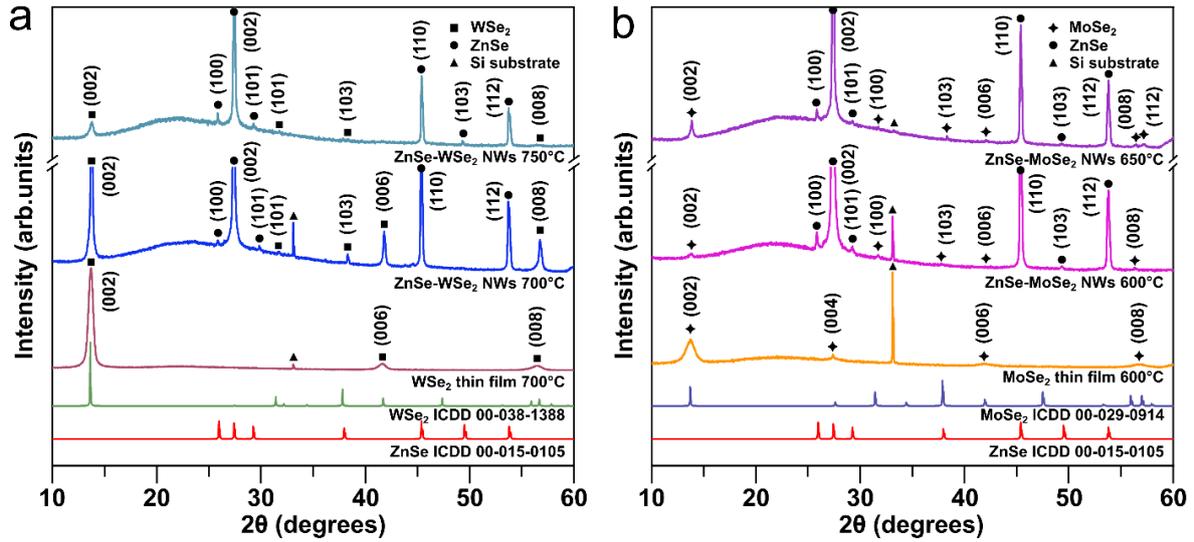

**Figure 2.** X-ray diffraction patterns of (a) ZnSe-WSe$_2$ core-shell NW, WSe$_2$ thin films and (b) ZnSe-MoSe$_2$ core-shell NWs, MoSe$_2$ thin films, all prepared on Si(100)/SiO$_2$ substrates and selenized at different temperatures, as well as of pure ZnSe NWs. The corresponding ICDD reference patterns for each phase are included. The symbols used to label the materials are as follows: square - WSe$_2$, circle - ZnSe, triangle - Si substrate, and cross - MoSe$_2$.

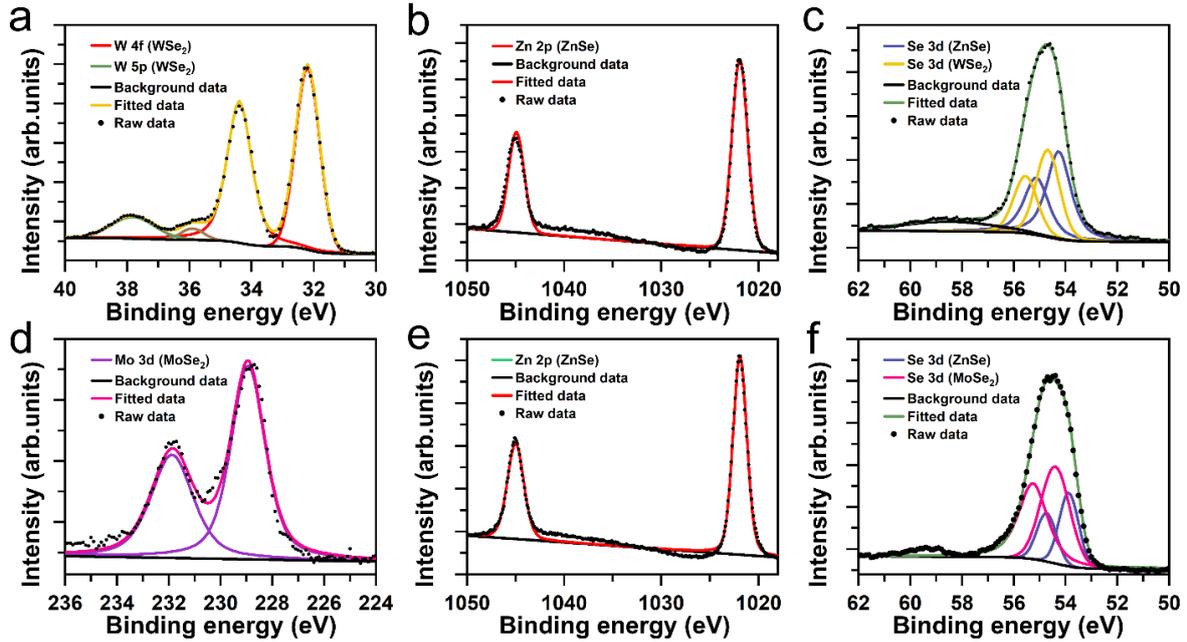

**Figure 3.** XPS spectra of (a,b,c) ZnSe-WSe$_2$ core-shell NWs selenized at 700°C and (d,e,f) ZnSe-MoSe$_2$ core-shell NWs selenized at 600°C, showing the high-resolution scans of the detected elements.

Raman spectroscopy of ZnSe-WSe$_2$ and ZnSe-MoSe$_2$ core-shell NWs, as well as reference WSe$_2$ and MoSe$_2$ thin films (Fig. 4), further verified the presence of MoSe$_2$ and WSe$_2$, confirming the formation of the intended core-shell heterostructures. The Raman spectra of both core-shell NWs exhibited peaks at ~139 cm$^{-1}$ (2TA), ~204 cm$^{-1}$ (TO), and ~251 cm$^{-1}$ (1LO), which correspond to ZnSe [37], indicating that the ZnSe core remained structurally intact after shell formation. For ZnSe-WSe$_2$ NWs and the WSe$_2$ thin film, additional peaks were observed at ~248 cm$^{-1}$ ($E^1_{2g}$), ~250 cm-1 ($A_{1g}$), and ~309 cm$^{-1}$ ($B^1_{2g}$), consistent with WSe$_2$ [38]. ZnSe-MoSe$_2$ NWs and the MoSe$_2$ thin film exhibited the characteristic peaks at ~241 cm$^{-1}$ ($A_{1g}$) and ~286 cm$^{-1}$ ($E^1_{2g}$), corresponding to MoSe$_2$ [38,39]. The absence of peaks associated with the precursor metal oxides suggests that the selenization process resulted in complete conversion.

Fig. 5 shows the PL spectra of ZnSe-WSe$_2$ and ZnSe-MoSe$_2$ core-shell NWs, as well as pure ZnSe NWs and corresponding WSe$_2$ and MoSe$_2$ thin films, the PL intensity is shown in arbitrary units and does not reflect the relative intensities between the different spectra. All intensities are normalized to 1, except for the thin films, which are normalized to 0.5 for the clarity. ZnSe NWs exhibit a broad red emission band centred around 620 nm, which is associated with Zn vacancies, this emission being consistent with the self-activated luminescence observed in II-VI compounds, originating from the recombination of shallow donor–acceptor pairs [40]. At 460 nm, a weaker emission feature is visible, corresponding to a ZnSe direct bandgap (2.7 eV) transition [41]. Notably, the same 460 nm peak is also observed in ZnSe-WSe$_2$ and ZnSe-MoSe$_2$ core-shell NWs. Both WSe$_2$ thin film and ZnSe-WSe$_2$ core-shell NWs (Fig.5(a)) exhibit a broad blue-green PL band centred at approximately 480 nm, which could be attributed to the so-called B' transition [42,43]. The B' transition corresponds to an exciton feature arising from valence band states, distinct from the primary B exciton. Similarly, in MoSe$_2$ thin film and ZnSe-MoSe$_2$ core-shell NWs spectra (Fig.5(b)) a wide PL band centred at 480 nm is observed in the blue-green region, which could be attributed to the B' transition similar as for WSe$_2$ [42,43].

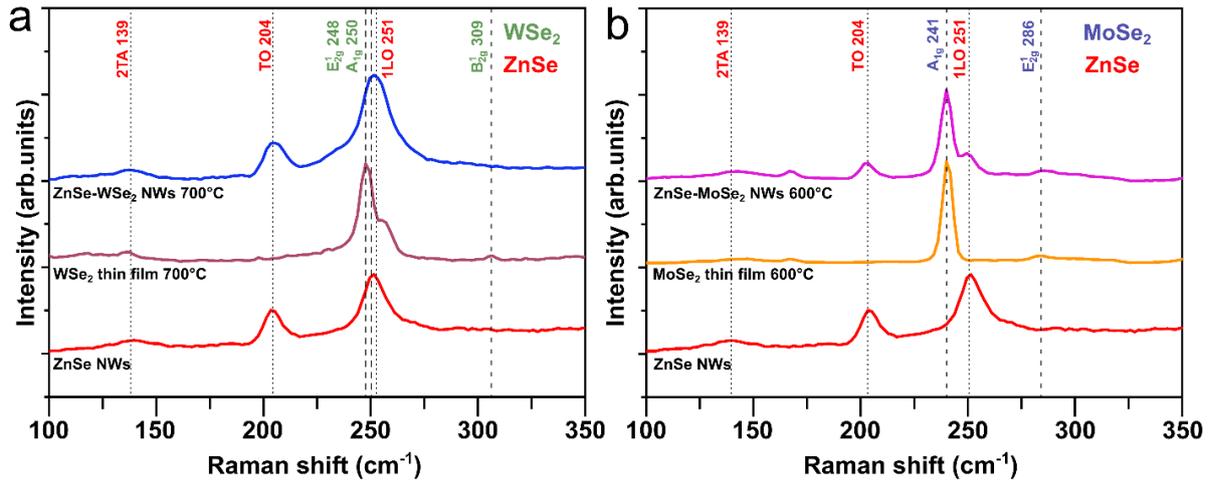

**Figure 4.** Micro-Raman spectra of (a) ZnSe-WSe$_2$ core-shell NWs selenized at 700°C, WSe$_2$ thin film selenized at 700°C and (b) ZnSe-MoSe$_2$ core-shell NWs selenized at 600°C, MoSe$_2$ thin film selenized at 600°, as well as of pure ZnSe NWs for reference, showing identified vibrational modes for each peak.

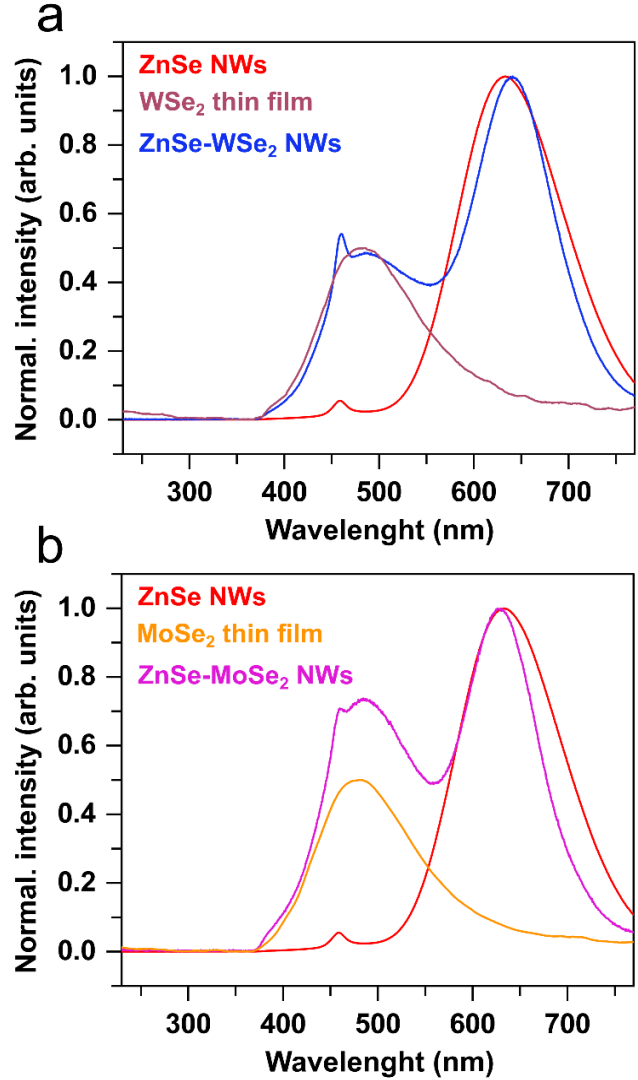

**Figure 5.** Photoluminescence spectra of (a) ZnSe-WSe$_2$ core-shell NWs selenized at 700°C, WSe$_2$ thin film selenized at 700°C and (b) ZnSe-MoSe$_2$ core-shell NWs selenized at 600° and MoSe$_2$ thin film selenized at 600°, as well as of pure ZnSe NWs.

## 4. Discussion

While qualitatively similar results were obtained with XRD, XPS, Raman, and PL spectroscopy for the samples synthesized at 600 °C and 650°C for ZnSe-MoSe$_2$, and 700°C and 750°C for ZnSe-WSe$_2$ core-shell, TEM study indicated a difference in the shell morphology. The formation of protruding TMD crystals at a higher selenization temperatures for both material combinations – ZnSe-WSe$_2$ and ZnSe-MoSe$_2$ – suggests a temperature-dependent transition from conformal shell growth to crystal nucleation and growth. We have observed similar trends for other TMD materials in our previous works [44–46], and we believe there are several mechanisms behind this effect, which ought to be considered when optimizing the core-shell NW morphology of other materials combinations. At relatively lower temperature, predominantly solid-state chemical reaction between the $H_2Se$ gas and/or selenium vapour and the sacrificial oxide coating occurs at the ZnSe-oxide interface, therefore, the final shell is smoother and its thickness is mainly determined by the initial precursor film thickness. Worth noting that the final shell is typically several times thinner than the initial sacrificial coating thickness due to some evaporation of the oxide, however, no selenide redeposition occurs meaning the process temperature is too low for efficient gas-phase reaction.

The symmetry of the lattices and lattice constant mismatch still plays an important role in achieving conformal shell, however, the restrictions for commensurate-like growth might be reduced in core-shell NWs and with van der Waals materials [47,48]. On the other hand, when the selenization temperature is increased, the vapour pressure of the sacrificial oxide film becomes significantly high, and the oxide starts to evaporate rapidly. The chemical reaction between the metal oxide and selenium then occurs in the gas phase, leading to the deposition of the formed TMD molecules onto the NW surface from the gas phase and growth of crystalline islands. Such growth process follows the typical CVD diffusion-enabled nucleation and island growth principles. Therefore, to obtain the desired TMD morphology on NWs, the main parameters to consider are temperature, precursor film vapour pressure and thickness, assuming the selenium vapour partial pressure is constant and the oxide coating is fully converted (optimized process duration). While at lower selenization temperature the shell is more uniform and its thickness is closely related to the initial sacrificial layer thickness although typically several times thinner, at higher temperature the interplay between the precursor film evaporation and chalcogenide deposition needs to be examined. Uniform and highly-crystalline shell would be needed for applications in electronics and optoelectronics, while the large surface area of the protruding TMD crystals would give significant benefits in catalytic applications. To demonstrate the original functional properties of our fabricated core-shell NW, photoelectric measurements should be performed on single-nanowire devices in order to study their feasibility for optoelectronic applications[49], while experimental and theoretical study of their photocatalytic properties should be performed to elaborate their applicability in hydrogen evolution reaction or similar photocatalysis applications [50,51].

We believe the observed trend can be applied to other TMDs and layered van der Waals materials, however, depending on the corresponding metal oxide vapour pressure, a sacrificial metal film might lead to better results. For example, TEM images of ZnSe-ReSe$_2$ core-shell NWs (see Fig.S7), grown by selenization of sacrificial Re metal sputtered on ZnSe NW, indicate formation of a smooth shell, while it was only possible to obtain protruding crystals when using rhenium oxide precursor due to its low vapour pressure. However, it must be noted that for some transition metals both oxide and metal has very low vapour pressure at temperatures in which the core material can survive without degradation, therefore, the conversion regime for such shell materials might be limited just to the solid-state reaction.

## 5. Conclusions

Successful synthesis of ZnSe-WSe$_2$ and ZnSe-MoSe$_2$ core-shell NWs through selenization of a sputter-coated oxide precursor on ZnSe NWs has been demonstrated. XRD confirmed the phase composition of the resulting structures, showing the presence of wurtzite ZnSe, MoSe$_2$, and WSe$_2$ phases with no detectable peaks from precursor oxides, indicating complete conversion. The presence of the compounds was also confirmed by XPS and Raman spectroscopy. Optical properties of the heterostructures were studied by PL spectroscopy. TEM measurements confirmed the formation of conformal, layered TMD shells at 600 °C and 700 °C for WSe$_2$ and MoSe$_2$, respectively, while at higher selenization temperatures (650 °C and 750 °C, respectively), shell formation was not uniform, because of the growth of protruding

TMD crystals on the NW surface, indicating a temperature-dependent shift in the growth mechanism. Possibly, at lower temperature, predominantly solid-state chemical reaction between the $H_2Se$ gas / Se vapour and the sacrificial oxide coating occurs, leading to smoother shell with a thickness mainly determined by the initial precursor film thickness although typically several times thinner; while at higher temperature the interplay between the precursor film evaporation, gas-phase chemical reaction, and chalcogenide deposition needs to be considered. The observed trends could be extended to other semiconductor NW and TMD core-shell heterostructure growth, offering promising avenues for advanced nanoscale applications.

**Supplementary Materials:** Supporting information is available and contains characterization data for pure ZnSe NWs reference sample, XPS spectra for the ZnSe, $WSe_2$ and $MoSe_2$ reference samples, and TEM images of ZnSe-$ReSe_2$ core-shell NWs prepared via conversion of a pre-deposited 15 nm thick Re metal coating.

**Funding:** This research was funded by the Latvian Council of Science project No. lzp-2022/1-0311.

**Acknowledgements:** The authors are grateful to Kevon Kadiwala for his assistance with the development of the growth process.

**Author Contributions:** Conceptualization, E.B. and B.P.; Methodology, E.B. and B.P.; Validation, L.D., L.K., E.B.; Investigation, L.D., L.K., V.V., K.L., A.Z., E.D., J.G., E.B. and B.P.; Writing – Original Draft Preparation, L.D., E.B.; Writing – Review & Editing, L.D., E.B., B.P., K.L.; Visualization, L.D.; Supervision, E.B.; Project administration, E.B.

**Institutional Review Board Statement:** Not applicable.

**Informed Consent Statement:** Not applicable.

**Data Availability Statement:** The data supporting this study's findings are available from the corresponding author upon reasonable request.

**Conflicts of Interest:** The authors declare no conflict of interest.